\documentclass[conference]{IEEEtran}

 \IEEEoverridecommandlockouts
\usepackage{cite}
\usepackage{amsmath,amssymb,amsfonts}
\usepackage{algorithmic}
\usepackage{graphicx}
\graphicspath{ {./images/} }
\usepackage{textcomp}
\usepackage{tikz}
\usepackage{multirow}
\usepackage{graphicx}
\usetikzlibrary{shapes.geometric, arrows}
\usepackage{xcolor}
\def\BibTeX{{\rm B\kern-.05em{\sc i\kern-.025em b}\kern-.08em
    T\kern-.1667em\lower.7ex\hbox{E}\kern-.125emX}}
\begin{document}

\title{FL-DECO-BC: A Privacy-Preserving, Provably Secure, and Provenance-Preserving Federated Learning Framework with Decentralized Oracles on Blockchain for VANETs\\
}

\author{Sathwik Narkedimilli, Dr. Pavan Kumar C, Rayachoti Arun Kumar, N. V. Saran Kumar, Ramapathruni Praneeth Reddy\\
\textit{Department of Computer Science, Indian Institute of Information Technology (IIIT) Dharwad, Dharwad, India}}

\maketitle
 
\begin{abstract}
Vehicular Ad-Hoc Networks (VANETs) hold immense potential for improving traffic safety and efficiency. However, traditional centralized approaches for machine learning in VANETs raise concerns about data privacy and security. Federated Learning (FL) offers a solution that enables collaborative model training without sharing raw data. This paper proposes FL-DECO-BC, a novel privacy-preserving, provably secure, and provenance-preserving federated learning framework specifically designed for VANETs. FL-DECO-BC leverages decentralized oracles on blockchain to securely access external data sources while ensuring data privacy through advanced techniques. The framework guarantees provable security through cryptographic primitives and formal verification methods. Furthermore, FL-DECO-BC incorporates a provenance-preserving design to track data origin and history, fostering trust and accountability. This combination of features empowers VANETs with secure and privacy-conscious machine-learning capabilities, paving the way for advanced traffic management and safety applications.
\end{abstract}


\section{Introduction}
Advanced road navigation techniques called Intelligent Transportation Systems (ITS)~\cite{al2021security} are the way to the future. They utilize Vehicle Ad-hoc Networks~\cite{elsagheer2021intelligent} to promote real-time communication between vehicles. This results in improved safety and efficiency for traffic. Artificial Intelligence also brings significant opportunities to VANETs. For example, one can imagine a vehicle training a machine learning model that identifies areas where accidents happen from the data collected by multiple vehicles, potentially preventing accidents, as explored in research on federated learning for cooperative driving ~\cite{electronics12040894}. However, while such measures could be widely implemented, several challenges remain on the way, as discussed in ~\cite{article123} .\\

Federated Learning is a promising infrastructure for collaborative model training in VANETs, but it also presents several challenges that affect data privacy, security, and provenance preservation. In VANETs, data privacy is a major concern since sensitive data, such as vehicle speed, location, and other vehicle-related information, is collected. Leaks can lead to user identification and unwanted tracking, creating user safety vulnerabilities. Security is essential to safeguard the training process since malicious users can publish false or altered data that leads to an untrusted or malicious model. Provenance is also essential because the data source and data history are not accessible, making it challenging to evaluate the training data's bias or the data itself.
\\

Although current solutions aim to ensure secure learning (FL) in vehicular ad hoc networks (VANETs), they may face limitations. Enter the FL-DECO-BC framework, an alternative that not only solves these challenges but also preserves the original data, a notable modification of which FL-DECO-BC works through a variety of mechanisms:\\

\begin{enumerate}
    \item \textit{Local Model Training:} Vehicles equipped with on-board units (OBUs) and roadside units (RSUs) use their collected data to train their individual models.
    \item \textit{Secure Weight Upload:} Instead of raw data, trained model loads are securely uploaded to the VANET blockchain network via decentralized oracles ~\cite{beniiche2020study}. These oracles ensure the integrity and validity of the data before uploading it.
    \item \textit{Secure Aggregation with Verifiable Computation:}  computation statements take weights from blockchain and perform secure multi-party computation (SMPC) to aggregate model updates ~\cite{liu2022privacypreserving} Notably, FL-DECO-BC brings a unique feature where multiple oracles are independent enter SMPC, increases security. Furthermore, the testimony presented by the speakers attests to the correctness of the calculations, further reinforcing the system's integrity.
    \item \textit{Model Deployment and Utilization:}  The aggregate model is securely stored on the blockchain for VANET applications.\\
\end{enumerate} 
FL-DECO-BC is a secure solution for Federated Learning (FL) in Vehicular Ad-hoc Networks (VANETs). It uses decentralized oracles and verifiable computation, and can leverage multiple computation oracles to enhance security. FL-DECO-BC stores model updates and their associated data provenance on the blockchain, allowing us to track the origin and history of the data used in training. This is very important for ensuring transparency and accountability, as it helps to identify potential biases or errors within the model.\\ \\
The following sections examine the study's specifics in more detail. Section 2, entitled Preliminaries, establishes the foundational concepts. Section 3, Literature Review, surveys existing research on the topic. The main body of the paper is in Section 4, Proposed Models, where the new models presented are introduced. Section 5 of the evaluation of the proposed model examines its effectiveness. Finally, Section 6, Conclusion, summarizes the main findings and implications.

\section{Preliminaries}

\subsection{Vehicular Ad-Hoc Networks (VANETs)}

Vehicular Ad-Hoc Networks (VANETs) are a revolutionary technological development in intelligent transportation systems ~\cite{al2021security}. This is a network that is not centralized, and it allows vehicles to communicate directly with one another, V2V, or the roadside infrastructure V2I ~\cite{al2021security}, without any need for a central access point. This network is revolutionary as it changes the face of travel and avails applications that enhance

\begin{enumerate}
 \item \textit {Traffic Safety:}Drivers can actually avoid any forthcoming road traffic accidents by responding promptly to real-time road closures, accident news, and other dangerous situations available through vehicle-to-vehicle communications exchange ~\cite{al2021security}.
\item \textit{Traffic Efficiency:}Traffic congestion data and the best pathways can be swapped, thus optimizing traffic flow and reducing fuel consumption and travel times, for example [1].
\item \textit{Driver Comfort:}Distance makes a huge difference regarding driver comfort. The information that could help when one is driving is given in relation to the distance covered from the source to the destination. These may include areas such as parking, fuel stations, and weather ~\cite{al2021security}.\\
\end{enumerate} 

However, VANETs also introduce unique challenges:
\begin{enumerate}
\item \textit{Data Privacy:}VANET data often contains sensitive information about drivers' locations, behavior, and surrounding environments. Protecting this data from unauthorized access or misuse is paramount ~\cite{al2021security} ~\cite{elsagheer2021intelligent}.
\item \textit{Data Provenance:} Distance makes all the difference Regarding driver comfort. The nature of information that can be helpful during driving is provided about the number of miles traveled from point A to point B. This may include aspects like parking lots, gas stations, and weather. VANET data usually contains personal details about drivers’ whereabouts, actions, and immediate surroundings. Such information must not be compromised or unlawfully accessed. There is a need for officials handling safety emergencies to establish where the data came from so that it can serve its intended purposes. It also helps to ensure responsibility and avoid spreading rumors across the network. For example, some crooks could inject fake accidents and hazardous messages, disrupting and endangering security precautions ~\cite{article123}.
\end{enumerate}

\subsection{Decentralized Oracles (DECO)}
Decentralized oracles ~\cite{beniiche2020study}(DECO) are blockchain-intermediated networks that work as reliable mediators between external data and smart contracts on the blockchain by transforming the former into the latter. This fills in the gap between data originations of real-world that is risky and blockchain that is safe but isolated from other systems.\\
{Mechanisms for Decentralized Oracles~\cite{10230033}: }
\begin{enumerate}
    \item \textit{Stake-Based Reputation Systems ~\cite{chainlink}:} In an Oracle network, nodes usually stake a specific quantity of cryptocurrency/tokens. This stake's collateral motivates them to supply accurate and trustworthy data. Penalties and stake loss may result from submitting incorrect data.
    \item \textit{Consensus Procedures:} Decentralized oracles, like blockchains, use consensus procedures to guarantee that the authenticity of the data that is fetched is agreed upon. Different projects use different algorithms, including Proof-of-Stake (PoS) or Byzantine Fault Tolerance (BFT), to reach a consensus.
    \item \textit{Data Feeds and Aggregation:} Web servers, APIs, and even Internet of Things devices are just a few of the data sources that Oracles may access. After being recovered, the data may be preprocessed, combined, and added to the blockchain.\\
\end{enumerate}
{The advantages of using decentralized oracles in VANETs are:}
\begin{itemize}
    \item \textit{Data Privacy:} Oracles protect individual driver information by anonymizing datasets while making it possible to use critical group insights.
    \item \textit{Provenance Protection ~\cite{chainlink}:} Origins and history of decentralized oracles using the immutability of blockchain can produce an indelible record.
\end{itemize}

\subsection{Federated Learning(FL)}
Federated learning ~\cite{pmlr-v54-mcmahan17a} has emerged as a revolutionary approach to machine learning in scenarios where data privacy is a major concern. Unlike traditional methods that centralize data for training, FL enables collaborative learning on distributed devices while keeping data private. The steps include :

\begin{enumerate}
    \item \textit{Model Distribution:} A central server distributes a pre-trained machine learning model to participating devices (e.g., smartphones in VANETs).
    \item \textit {Local Training:} Each device trains the model locally using its own private data. This ensures data never leaves the device itself.
    \item \textit{Model Updates:} After local training, devices only share the model updates (weight changes and biases) with the central server. These updates don't reveal the original data.
    \item \textit{Global Model Update:} The central server aggregates the model updates from all devices and uses them to improve the global model.
    \item \textit{Iteration:} This process of distributing the updated model, local training, and model update aggregation is repeated iteratively until the global model achieves the desired performance.\\
\end{enumerate}

 {Benefits of Federated Learning:}

\begin{itemize}

    \item \textit{Privacy Protection:} FL ensures privacy by storing data on individual devices, thus minimizing the risks associated with centralized data storage ~\cite{electronics12040894} ~\cite{pmlr-v54-mcmahan17a}.
    \item \textit{Data Security:} Even when model updates are shared with the server, they are anonymized, reducing susceptibility to attacks targeting training data.
    \item \textit{Scalability:} FL harnesses the computational capabilities of numerous devices, facilitating the training of intricate models across distributed datasets.
    \item \textit{Data control:} Users maintain ownership of their data and have the discretion to decide whether to engage in the learning process or not.
\end{itemize}

\subsection{Blockchain}
Blockchain technology ~\cite{article8957639} underpins the foundation of many revolutionary applications. It offers a secure, transparent, and distributed way to record and manages data. Here's a breakdown of its core principles:

\subsection*{Core Principles:}

\begin{enumerate}
    \item \textit{Distributed Ledger:} A blockchain is a distributed ledger, meaning a replicated database shared across a network of computers (nodes). This eliminates the need for a central authority to control the data.
    \item \textit{Immutability:} Transactions on a blockchain are cryptographically linked together in a chain of blocks. Each block contains data (e.g., financial transactions, data records) and a unique hash, a mathematical fingerprint. Any attempt to tamper with a block would alter its hash, making the change readily detectable by all participants in the network. This ensures the immutability and tamper-proof nature of the data stored on the blockchain.
    \item \textit{Consensus Mechanisms:} To maintain consistency across the distributed network, blockchains employ consensus mechanisms. These algorithms ensure that all nodes agree on the validity of transactions and the current state of the ledger. Popular consensus mechanisms include Proof-of-Work (PoW) ~\cite{article8957639} and Proof-of-Stake (PoS) ~\cite{article8957639}.
\end{enumerate}

\subsection*{Benefits of Blockchain:}

\begin{itemize}

    \item \textit{Security:} Blockchains utilize distributed architecture and cryptographic methods, rendering them highly immune to cyber threats and tampering with data ~\cite{electronics12030546}.
    \item \textit{Transparency:} All network members can access the complete transaction history, promoting trust and responsibility ~\cite{electronics12030546}.
    \item \textit{Efficiency:} By removing intermediaries and enabling secure, automated transactions, blockchains enhance process efficiency ~\cite{electronics12030546}.
\end{itemize}

\subsection{Cryptographic Primitives}
a. Zero-Knowledge Proofs:

Imagine a scenario where you want to convince someone you are above 21 years old to enter a bar without revealing your actual age. Zero-knowledge proofs ~\cite{articleOverviewandApplications} provide a cryptographic solution to this problem. They allow one party (the prover) to convince another party (the verifier) that they possess a certain piece of information (e.g., being over 21) without actually revealing the information itself.

Here's a simplified breakdown of how zero-knowledge proofs work:

\begin{enumerate}
    \item \textit{Setup:} The prover and verifier agree on a common protocol and the specific statement to be proven (e.g., `the prover's age is greater than 21`).
    \item \textit {Interactive Proof:} The prover interacts with the verifier by sending mathematical proofs and challenges. These proofs are designed so that only someone with knowledge of the secret information (being over 21) can generate them.
    \item \textit{Verification:} The verifier checks the validity of the proofs received from the prover. If all the proofs are valid, the verifier is convinced that the prover possesses the secret information with high certainty without ever learning the actual value (age in this example).
\end{enumerate}

{Applications:}

Zero-knowledge proofs extend beyond age verification and find utility in various areas, including:

\begin{itemize}
    \item \textit {Digital Signatures}: Individuals can demonstrate ownership of a private key linked to a public key without divulging the private key during the digital signing.
    \item  \textit {Identity Management}: Users can verify their membership in a specific group without revealing any personal details.\\
\end{itemize}

b. Secure Multi-Party Computation (SMPC):

Imagine a scenario where two parties want to calculate the sum of their private numbers without revealing those numbers to each other. Secure Multi-Party Computation (SMPC) ~\cite{ZHAO2019357} allows multiple parties to jointly compute a function on their private inputs while keeping those inputs confidential.

Here's a high-level overview of SMPC:

\begin{enumerate}
    \item \textit {Function Definition:} The parties agree on the function they want to compute collaboratively (e.g., addition in our example).
    \item \textit {Secret Sharing:} Each party splits their private input into secret shares and distributes them to all other participants. These shares are designed such that no single party can determine the original input from their share alone.
    \item \textit {Distributed Computation:} The parties perform a series of mathematical operations on their secret shares according to the agreed-upon protocol.
    \item \textit {Secret Reconstruction:} Once the computation is complete, each party can combine the shares they received from others to reconstruct the final result (the sum in our example) without ever revealing their original input.
\end{enumerate}

\section{Literature Review}
The potential of Federated Learning (FL) for Vehicular Ad-hoc Networks (VANETs) is a rapidly growing area of research. Several recent studies explore FL applications in VANETs for various purposes, emphasizing the benefits of privacy preservation and distributed processing. For instance, \textit{`Intrusion Detection in VANET Data Streams Using Federated Learning for Smart City Environments (2023)`} ~\cite{electronics12040894} investigates its use for anomaly detection in data streams. This approach allows vehicles to collaboratively train a model to identify unusual patterns in the data without sharing raw data, protecting privacy and reducing network load.\\

Furthermore, FL opens doors to a wide range of applications within Intelligent Transportation Systems (ITS), including VANETs. A broader perspective is provided in \textit{`A Survey on Federated Learning for Intelligent Transportation Systems: Challenges and Opportunities (2022)`}~\cite{zhang2024survey}. This survey highlights the challenges and opportunities related to security, privacy, and communication efficiency. \\

Specific applications explored in other studies include optimizing traffic flow while maintaining data privacy (\textit{`A Novel Reinforcement Learning-Based Cooperative Traffic Signal System Through Max-Pressure Control`}) ~\cite{9392372}, developing energy-efficient routing protocols (\textit{`Federated Learning in Vehicular Networks: Opportunities and Solutions`})~\cite{articlePosner}, and collaboratively detecting drowsy drivers to promote road safety (\textit{`Privacy enabled driver behavior analysis in heterogeneous IoV using federated learning`})~\cite{CHHABRA2023105881}. Additionally, \textit{`Fed-NTP: A Federated Learning Algorithm for Network Traffic Prediction in VANET`} ~\cite{9950054} explores predicting both aspects for more comprehensive traffic management.\\

In February 2024, Routis et al. published a research paper proposing a Federated Learning (FL) framework for Vehicle Ad-Hoc Networks (VANETs) that prioritizes privacy and security. The paper, titled `A Secure and Privacy Preserved Infrastructure for VANETs based on Federated Learning with Local Differential Privacy` ~\cite{BATOOL2024119717}, suggests that vehicles add noise to their data before sharing it with a central coordinator to protect their privacy. This approach provides a balance between privacy and accuracy but requires careful consideration of the computational limits of vehicles. The research offers insights into infrastructure design principles that prioritize privacy while acknowledging practical constraints, paving the way for more secure and efficient FL applications in VANETs. These studies demonstrate the diverse applications of FL that can revolutionize VANETs and contribute to safer, more efficient transportation systems.\\

While the discussed studies showcase the potential of Federated Learning (FL) for VANETs, significant security concerns must be addressed to ensure user trust and wider adoption. One major threat is ensuring data provenance, which refers to the origin and history of data. Without proper data provenance, malicious actors could inject false information into the system, impacting the accuracy and reliability of the FL model. However, current methods for tracing data provenance often struggle to do so without compromising user privacy. This creates a tension between robust data verification and user anonymity.\\

Furthermore, FL inherently makes it difficult to pinpoint who contributed what data and how it was used in the model training process. This lack of accountability can be a major hurdle for regulatory compliance and user trust. Without clear accountability mechanisms, enforcing regulations or holding bad actors responsible for manipulating data or compromising the system is challenging. Researchers are actively seeking solutions to these challenges, and future advancements in FL security will be crucial for establishing trust and enabling widespread adoption of FL in VANETs.\\

Another challenge is Sybil attacks. In these attacks, malicious actors impersonate multiple vehicles to manipulate the training data, potentially skewing the results and compromising the integrity of the FL model. While crucial for protecting user data, existing privacy-preserving techniques might not be sufficient to detect and prevent Sybil attacks.

\section{Proposed Model}

\subsection{The FL-DECO-BC Workflow}
The core of the FL-DECO-BC framework lies in its four-step process that leverages decentralized oracles, federated learning, and blockchain technology. This section will delve into each step in detail, explaining how local model training on devices, secure weight storage with decentralized oracles, privacy-preserving aggregation using SMPC, and blockchain-based model storage and utilization work together to achieve a secure, provenance-preserving, and privacy-preserving approach to federated learning for VANETs. The steps include : 

\begin{enumerate}
\item \textit {Local Model Training}: On-board units (OBUs) and Road Side Units (RSUs) within the VANET collect data relevant to the training objective. This data could include vehicle speed, location information, or sensor readings related to the environment. Each OBU and RSU trains a local machine-learning model on its collected data. This local training protects raw data privacy, as the data never leaves individual devices. The specific machine learning algorithm chosen for local training depends on the application and the desired outcome of the global model.\\

\item \textit {Secure Model Weight Storage with Decentralized Oracles}: After the local training, the On-Board Units (OBUs) and Road-Side Units (RSUs) transfer the trained model weights to the VANET blockchain network. However, to maintain data privacy, the weights are not uploaded directly. Instead, FL-DECO-BC uses decentralized oracles as intermediaries between the devices and the blockchain. These oracles can be implemented as smart contracts or trusted entities within the network. The model weights are encrypted and sent to the decentralized oracles. The specific consensus mechanism used by the decentralized oracles ensures that only authorized entities can contribute to storing the weights on the blockchain. This mechanism helps prevent unauthorized manipulation of the training data.\\

\item \textit {Secure Aggregation with SMPC and Verification}: The encrypted model weights are uploaded to the blockchain network through decentralized oracles. Secure Multi-Party Computation (SMPC) takes the model weights from the blockchain to generate a global model. Multiple computation oracles within the network collaborate in the SMPC process. SMPC allows oracles to compute the aggregate model weights on encrypted data without decrypting each device's individual contributions. This protects the privacy of the data used to train local models. After the SMPC computation, the oracles generate proofs that mathematically demonstrate the aggregated model's correctness.\\

\item \textit {Blockchain-Based Model Storage and Utilization}: The blockchain verifies the proofs (Zero Knowledge proofs) submitted by the computation oracles. This verification ensures that the aggregated model weights were correctly computed based on the uploaded individual weights. If the verification is successful, the final aggregated model is stored immutably on the blockchain network. The immutability of the blockchain ensures that the model cannot be tampered with after it's stored. Authorized participants within the VANET can then access and utilize the stored global model for various applications, such as traffic prediction, congestion control, or safety hazard warnings.\\
\end{enumerate}

\begin{figure}[h]
  \centering
  \includegraphics[width=8cm, height=8cm]{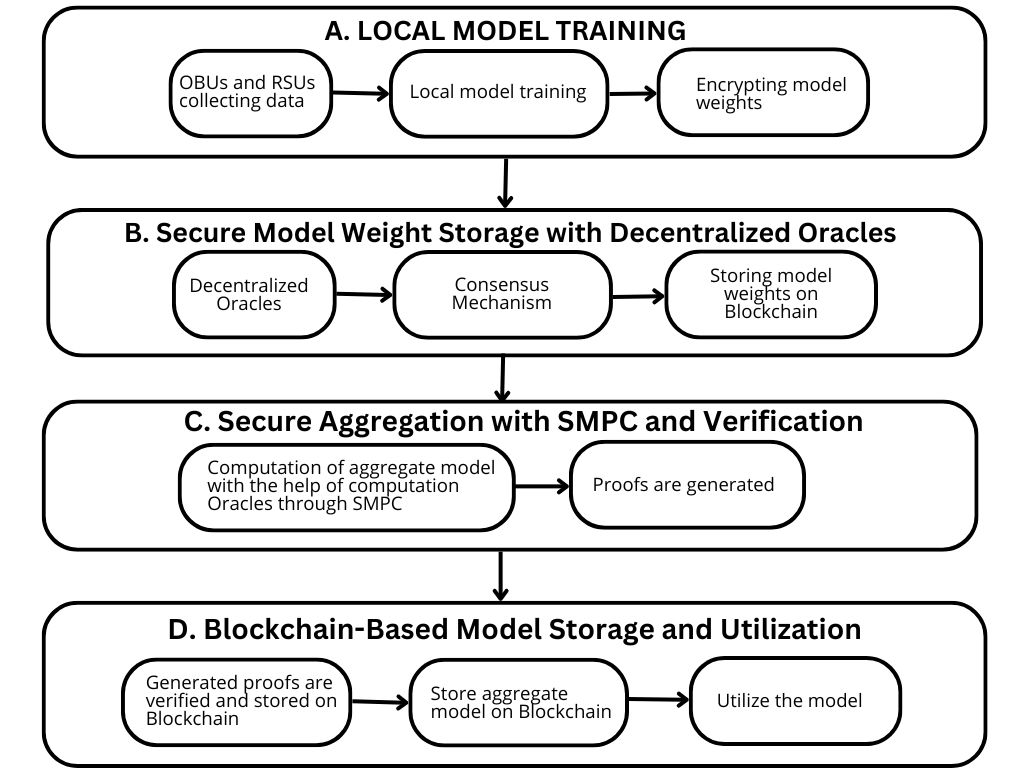}  
  \caption{Flow Diagram of FL-DECO-BC Framework}  
\end{figure}

\subsection{Assumptions for the FL-DECO-BC Framework}
The effectiveness of the proposed FL-DECO-BC framework relies on several critical assumptions. These assumptions ensure the proper functioning of each step within the framework and contribute to the overall security and efficiency of the system.\\

\textit{Data Quality and Resource Constraints at Edge Devices:} The framework assumes that the data collected by On-Board Units (OBUs) and Roadside Units (RSUs) is of high quality. This means the data should be accurate, reliable, and contain minimal noise, errors, or inconsistencies. This high-quality data is essential for training effective local models at the network's edge. OBUs and RSUs are assumed to possess sufficient computational resources to train these local models efficiently. The training process should not significantly impact the core functionalities of these devices, such as real-time communication or sensor data collection.\\

\textit{Blockchain Network Properties and Secure Communication:} The FL-DECO-BC framework leverages a secure and scalable VANET blockchain network. This network needs to efficiently handle the storage and retrieval of model weights generated during local model training. Furthermore, the network's scalability is crucial to accommodate the growing number of participants and the increasing volume of data generated as the system expands. Secure communication channels are also essential between the blockchain network and the computation oracles involved in the Secure Multi-Party Computation (SMPC) process. These channels ensure the confidentiality and integrity of data during the computation, protecting it from unauthorized access or manipulation.\\

\textit{Verification and Trust Assumptions:} The framework relies on the efficient verification of proofs generated by computation oracles. These proofs guarantee the correctness of the aggregated model computed using SMPC. The underlying blockchain network is assumed to be able to verify these proofs efficiently without significant computational overhead. Additionally, the overall trustworthiness of the VANET blockchain network is crucial. The network needs to be secure against unauthorized access or manipulation of stored data, particularly the aggregated model.\\

While the FL-DECO-BC framework incorporates security measures, the number of malicious actors within the network is assumed to be limited. Finally, stable network connectivity between OBUs, RSUs, oracles, and the blockchain network is essential for efficient communication and data exchange throughout the collaborative learning process.\\

\subsection{BAN Logic Analysis for FL-DECO-BC Framework}

BAN logic (Burrows-Abadi-Needham logic) ~\cite{inproceedings} is a formal method for analyzing security protocols, especially those involving authentication. It assumes attackers can eavesdrop but not alter messages.BAN logic helps find protocol weaknesses by showing if these beliefs can't be guaranteed.  While not foolproof, it's a valuable tool for building secure communication.

While BAN logic isn't directly applicable to Step 1 (local model training) and Step 4 (Blockchain-Based Model Storage and Utilization) due to the absence of message exchange, we can analyze steps involving communication between entities:\\

{\textit{BAN Logic for Step 2 (Secure Model Weight Storage with Decentralized Oracles):}}

Participants:
\begin{itemize}
    \item $S$: OBU or RSU (sender of model weights)
    \item $O$: Decentralized oracle network (receiver)
    \item $B$: Blockchain network (receiver)
\end{itemize}

Messages:
\begin{itemize}
    \item $M$: $\{Encrypted Model Weights\}_S$ (message sent by $S$)
\end{itemize}

BAN Logic Statements:
\begin{align*}
    &S \text{ believes } O \text{ fresh} \\
    &\text{ (freshness means } O \text{ hasn't been compromised)} \\
    &M \text{ only known to } S \text{ and } O \\
    &\text{ (due to encryption with a potentially secret key)} \\
    &\Rightarrow O \text{ receives } M \text{ from } S \text{ in fresh state}
\end{align*}

Assumptions:
\begin{itemize}
    \item The decentralized oracle network employs secure communication channels to prevent eavesdropping or tampering during message transmission.\\
\end{itemize}

{\textit{BAN Logic for Step 3 (Secure Aggregation with SMPC and Verification):}}

Participants:
\begin{itemize}
    \item $O_i$: Individual computation oracles (senders of proofs)
    \item $B$: Blockchain network (receiver)
\end{itemize}

Messages:
\begin{itemize}
    \item $M$: $\{Proof of Correctness\}$ (sent by each $O_i$)
\end{itemize}

BAN Logic Statements:
\begin{align*}
    &O_i \text{ believes } B \text{ fresh} \\
    &\text{ (freshness means } B \text{ hasn't been compromised)} \\
    &M \text{ only known to } O_i \text{ and } B \\
    &\text{ (due to the secure nature of SMPC)} \\
    &\Rightarrow B \text{ receives } M \text{ from } O_i \text{ in fresh state}
\end{align*}

Assumptions:
\begin{itemize}
    \item The SMPC implementation ensures only the correctness proof, not the individual model weights, is revealed to any participating oracle.
\end{itemize}

\section{Analysis of Proposed Framework}

FL-DECO-BC boasts a resilient architecture exhibiting resilience against a spectrum of data privacy and security attacks, ensuring the integrity and confidentiality of the collaborative learning process. To counteract data privacy attacks such as data poisoning, where malicious actors inject manipulated data to skew the model, FL-DECO-BC employs robust mechanisms. Firstly, the decentralized oracles play a crucial role in data validation, scrutinizing incoming data for anomalies and filtering out potentially malicious inputs before integration into the model, thereby fortifying the system against data manipulation attempts. Additionally, Secure Multi-Party Computation (SMPC) safeguards raw data privacy by facilitating the exchange of encrypted model updates during aggregation, preventing unauthorized access to sensitive information.\\

Moreover, FL-DECO-BC effectively addresses security attacks, particularly the vulnerability stemming from a single point of failure in traditional centralized models. By distributing tasks and computation across multiple entities via decentralized oracles, FL-DECO-BC eliminates a central target for potential attacks, bolstering its resilience against security breaches. Furthermore, the model implements Byzantine Fault Tolerance (BFT), which ensures correct computation even in the presence of faulty oracles by requiring a specific number of honest ones to reach consensus, thus fortifying the system against Byzantine failures that could disrupt its functionality. This comprehensive approach to security fortification underscores FL-DECO-BC's robustness and reliability in safeguarding against various forms of attacks in collaborative learning environments.\\

FL-DECO-BC is a robust framework that is dedicated to preserving the origin and evolutionary trajectory of model weights through a combination of blockchain technology and decentralized oracles. This ensures complete transparency and accountability in the system. The framework stores the final aggregated model and meticulously documents its complete provenance. Decentralized oracles ensure secure provenance preservation by validating model updates collaboratively before they are stored on the blockchain. This multi-party verification process mitigates the risk of manipulation and safeguards the integrity of the provenance record. FL-DECO-BC is a reliable and trustworthy framework for collaborative learning and model aggregation. 
\\

The following analysis details how our proposed framework holds up against various attacks and vulnerabilities:

\subsection{Anonymity}
FL-DECO-BC ensures anonymity throughout the federated learning process. Local models are trained on devices such as OBUs and RSUs without uploading the raw data. This prevents any information linking the data to a specific vehicle from being revealed. Additionally, decentralized oracles eliminate the need for a central server that could potentially access the data. Furthermore, these oracles only interact with the necessary model weights, further enhancing privacy. Secure multi-party computation (SMPC) is used to compute the aggregate model on the blockchain network, which only involves encrypted weights. This ensures that individual contributions from vehicles remain anonymous. This multi-layered approach effectively conceals the identity of vehicles participating in federated learning within the VANET.

\subsection{Non-traceability and Impersonation attacks}
FL-DECO-BC is a framework designed to combat non-traceability and impersonation attacks. It uses a combination of security mechanisms such as federated learning, decentralized oracles, blockchain technology, and secure multi-party computation (SMPC). In a federated learning setting, raw data never leaves on-board units (OBUs) and roadside units (RSUs), which makes it difficult to trace data back to a specific vehicle. Decentralized oracles eliminate a central point of vulnerability. Blockchain technology ensures data provenance, making tampering with or forging model weights impossible. SMPC adds another layer of defense by operating on encrypted weights that prevent oracles from deciphering individual contributions and thwarting impersonation attempts. By combining these four mechanisms, FL-DECO-BC protects against non-traceability and impersonation attacks.

\subsection{Message Modification Attack}
FL-DECO-BC employs a variety of measures to prevent replay and man-in-the-middle attacks effectively. One of these measures is the use of unique nonces attached to model weights during upload, which ensures that captured messages cannot be rebroadcasted by attackers. The decentralized nature of Distributed Oracles (DECOs) makes it difficult for attackers to intercept messages, while blockchain timestamps ensure that outdated messages are discarded. Moreover, Secure Multi-Party Computation (SMPC) works on encrypted data, making it incomprehensible to attackers even if they intercept it. All these defenses significantly reduce such attacks' effectiveness within the FL-DECO-BC framework.

\subsection{Replay attack and Man in the middle attack}
FL-DECO-BC adopts a multifaceted approach to thwart replay and man-in-the-middle attacks effectively. It uses freshness guarantees, such as unique nonces that are attached to model weights during upload, to prevent attackers from rebroadcasting captured messages. Additionally, the Distributed Oracles (DECOs) decentralized nature complicates interception attempts, while blockchain timestamps ensure outdated messages are discarded. Secure Multi-Party Computation (SMPC) operates on encrypted data, which makes intercepted information incomprehensible to attackers. All these collective defenses significantly reduce the effectiveness of such attacks within the FL-DECO-BC framework.\\

Additionally, Flexibility and Adaptability are inherent in FL-DECO-BC, as the framework can be tailored to various VANET applications with diverse data types and model requirements. By adjusting the model architecture and SMPC protocols, FL-DECO-BC can seamlessly adapt to scenarios such as traffic congestion prediction, accident detection, and collaborative route planning, ensuring its applicability across a spectrum of use cases while maintaining its robust security measures.\\
\clearpage
\renewcommand{\arraystretch}{1.5}

\begin{table}[t]
\centering

\begin{tabular}{| p{1.5cm}|p{3.8cm}||p{1cm}|p{1.5cm}|p{1.3cm}|p{1.5cm}|p{1.5cm}|p{1.5cm}|p{1.5cm}|  }

 \hline
 \multicolumn{9}{|c|}{A Comparative Table of FL-DECO-BC Framework and Other Frameworks} \\
 \hline
Groups&Attacks	& FedML-HE ~\cite{jin2023fedmlhe} &	 Differential Privacy + FL ~\cite{BATOOL2024119717}	& IDS + FL ~\cite{electronics12040894} & Privacy-Preserving-Group-Signatures + FL ~\cite{KANCHAN202393} &	Semi-Asynchronous-Hierarchical + FL ~\cite{chen2021semiasynchronous}&Contextual-Client-Selection + FL \cite{song2023v2xboosted} &	FL-DECO-BC (proposed)\\
 \hline
 
\multirow{13}{1.5cm}{Security}    &    DoS Attacks (Sybil) &	No & No & No & No  & Limited & Yes & Yes\\
&Spoofing Attacks	 & 	Limited & No 	 & No & No & 	Yes	 & Yes & Yes\\
&Tampering Attacks	 & Limited & 	Limited	 & Limited	 & No & Limited & Limited & Yes\\
&Replay Attacks	 & Yes	 & Limited  & 	Limited & - & 	Limited	 & Limited & Yes\\
&Byzantine Fault Tolerance Attack & - & - & - & - & - & No &Yes\\
&Backdoor attacks & No & No & - & - & 	- &  No &Yes\\
&Centralized Server Compromise & Limited & 	No	 & Yes & 	No & 	Yes & 	No&Yes\\
&Masquerading Attacks	 & 	Limited  & No	 & No & 	No  & 	Yes	 & Yes&Yes\\
&Front Running Attack & Limited & 	Yes	 & No & 	No & 	No & 	- &Yes\\
&Message Modification&Limited	 & Limited	 & Limited & 	Yes & 	Limited & Yes& Yes\\
&Man in the Middle Attack&Limited	 & Limited	 & Limited& 	- & 	Limited & Limited& Yes\\
&Model Inversion & 	Limited & 	- & 	- & 	- & 	- & Limited& Yes\\

 \hline
\multirow{5}{1.5cm}{Privacy Preserving} & Eavesdropping	 & Yes & 	Limited	 & Yes	 & Yes 	 & Yes	& Yes &Yes\\
&Location Pinpointing & 	Limited	 & No  & Limited & Yes & Limited  & Limited	&Yes\\
&Anonymity&  Yes & Yes 	 & Yes & - & Yes & Yes & Yes\\
&Non-traceability and Impersonation Attacks&Yes & Yes 	 & Yes & - & Yes & Yes & Yes\\
&Traffic Analysis	 & Limited	 & Limited	 & Limited & 	Yes & 	Limited & Limited& Yes\\
\hline
\multirow{3}{1.5cm}{Provenance Preserving} 
&Data Poisoning & 	Limited & 	Limited	 & -	 & -& 	Limited& 	Limited& Yes \\
&Side Channel Attack&No&Yes&No&No&-&-&No\\
&Colluison Attack &Limited&Yes&No&Yes&-&No&Yes\\

 \hline
\end{tabular}

\end{table}

\section{Conclusion}
This FL-DECO-BC is a new federated learning framework that addresses privacy, security, and data provenance preservation challenges in Vehicular ad-hoc networks (VANETs). This framework uses decentralized oracles on blockchain technology to facilitate secure access to external data sources that are crucial for VANET applications. Advanced techniques ensure data privacy, while cryptographic primitives and formal verification methods provide provable security guarantees, which promote trust in the learning process. Moreover, FL-DECO-BC is designed to preserve data provenance, enabling the tracking of data origin and history and thus promoting accountability and fairness.

FL-DECO-BC enables secure and privacy-conscious machine learning in VANETs, leading to advanced traffic management applications that improve road safety and efficiency. Future research includes performance evaluations in real-world scenarios, exploring incentive mechanisms, and advancements in decentralized oracle functionalities to unlock the full potential of machine learning in VANETs.
\\
\\
\\
\\
\\
\\
\\
\\
\\
\\
\\
\\
\\
\\
\\
\\
\\
\\
\\
\\
\\
\\
\\
\\
\\
\\
\\
\\
\\
\\
\\
\\
\\
\\
\\
\bibliography{ref}
\bibliographystyle{IEEEtran}

\end{document}